# Whole Blood-Based Point-of-Care Assay for Single-Tier Lyme Disease Testing

## Authors

Mihye Lee,[1†] Rajesh Ghosh,[1†] Artem Goncharov,[2] Rui-Chian Tang,[1] Barath Palanisamy,[1] Gyeo-Re Han,[2] Adrian Anaya,[3] Elizabeth J. Horn,[4] Paul M. Arnaboldi,[5] Raymond J. Dattwyler,[5] Omai B. Garner,[6] Aydogan Ozcan,[1,2,7*] and Dino Di Carlo[1,7*]

## Affiliations

[1]Bioengineering Department, University of California, Los Angeles, CA, USA

[2]Electrical & Computer Engineering Department, University of California, Los Angeles, CA, USA

[3]Chemical Engineering Department, University of California, Los Angeles, CA, USA

[4]Lyme Disease Biobank, Portland, OR, USA

[5]Biopeptides, Corp., Ridgefield, CT, USA

[6]Department of Pathology and Laboratory Medicine, University of California, Los Angeles, CA, USA

[7]California NanoSystems Institute (CNSI), University of California, Los Angeles, CA, USA

†Contributed equally; *Corresponding author: ozcan@ucla.edu, dicarlo@ucla.edu

## Abstract

Rapid point-of-care diagnostics that operate directly on whole blood can accelerate clinical decision-making by bypassing complex sample preprocessing and centralized laboratory infrastructure. However, most serologic assays still require serum or plasma, limiting near-patient deployment. This challenge is particularly evident for Lyme disease, whose diagnosis relies on centralized two-tier serology. Here, we present a whole-blood multiplexed vertical flow assay (WB-xVFA) that performs single-tier Lyme serology directly from 50 µL of whole blood. The assay combines multilayer blood filtration with a peptide-patterned nitrocellulose membrane to separate cellular components and detect Lyme-associated IgM/IgG responses on a cartridge. Coupled with smartphone-based imaging and machine-learning analysis, the WB-xVFA provides results in 20 min. The assay remained robust across hematocrits up to 60% and sample volumes ranging from 25 to 100 µL. In blinded testing of 62 whole-blood measurements, it achieved ~97% sensitivity and specificity, demonstrating the feasibility of rapid, near-patient whole-blood Lyme serology.

## INTRODUCTION

Point-of-care (POC) diagnostics can shorten the interval between clinical presentation and treatment, yet serologic testing remains largely dependent on centralized laboratories because most assays are designed for serum or plasma obtained through off-device sample processing **(Fig. 1A)**[1–5]. This dependence is particularly consequential for time-sensitive infectious diseases, in which delays in diagnostic testing can postpone treatment[5–7]. Lyme disease exemplifies this challenge. As the most common vector-borne infection in North America and Europe, Lyme disease has an expanding geographic footprint and can progress to neurological, cardiac, or musculoskeletal complications when treatment is delayed[8–11]. Clinical diagnosis is often uncertain, particularly in the absence of the characteristic erythema migrans rash, necessitating laboratory diagnostic support[12,13]. Current testing relies primarily on centralized two-tier serology, which requires processed specimens, specialized laboratory infrastructure, and sequential testing procedures that can delay clinically actionable results[13,14]. Diagnostic sensitivity is also reduced during early infection, when pathogen-specific antibody responses are still evolving[15,16]. Together, these limitations underscore the need for rapid serologic testing that can operate directly from whole blood and provide clinically actionable results in near-patient settings.

Existing POC Lyme tests address only portions of this need. The Sofia 2 Lyme FIA, currently the only FDA-cleared POC test for Lyme disease, can use fingerstick whole blood; however, blood cells must first be removed using a separate capillary device before the processed sample is applied to the assay. In addition, the Sofia 2 functions only as a first-tier test, with reactive results requiring confirmatory second-tier testing in a centralized laboratory[17,18]. Despite advances in investigational POC Lyme serology, these platforms have largely remained dependent on serum or plasma[16,19–21]. Consequently, a rapid platform that combines direct whole-blood processing with multiplexed serologic sensing and single-tier diagnostic interpretation has not yet been demonstrated.

Direct use of whole blood introduces several coupled technical challenges. Red blood cells can obscure optical signals, contribute nonspecific background, and accumulate within porous assay materials, while hematocrit-dependent changes in cellular load, viscosity, and recoverable plasma volume can alter fluid transport and assay reproducibility[3,22,23]. These effects are particularly problematic for multiplexed serology, in which disease classification depends on measuring heterogeneous antibody responses distributed across multiple antigens rather than on a single biomarker. Conventional lateral-flow formats can incorporate blood-separation materials, but their predominantly one-dimensional flow path limits spatial multiplexing and couples sample transport directly to downstream sensing[24,25]. A rapid whole-blood serologic assay must therefore reconcile efficient cellular filtration with uniform delivery of antibodies to a multiplexed sensing region.

Vertical flow assay (VFA) architectures are well suited to these constraints. Their through-plane flow geometry enables sequential sample-processing layers to be stacked above a two-dimensional sensing membrane, supporting both upstream blood filtration and spatially multiplexed serologic detection[26–29]. Unlike lateral-flow formats, in which sample transport and detection occur predominantly along the same planar flow path, the stacked VFA configuration can distribute blood-cell removal across multiple layers before the soluble antibody fraction reaches the sensing array. This separation enables whole-blood filtration to be coupled with multiplexed antibody detection while preserving rapid transport and spatially resolved serologic measurements.

Here, we report a whole-blood multiplexed vertical flow assay (WB-xVFA) for rapid, single-tier Lyme disease serology directly from 50 µL of whole blood (**Fig. 1B-D**). The platform combines on-cartridge blood filtration with a peptide-patterned sensing array, portable smartphone imaging, and machine-learning interpretation to generate a diagnostic result within 20 min. We characterized WB-xVFA performance across physiologically relevant hematocrit levels and sample volumes and evaluated the assay using blinded clinical whole-blood specimens from the Lyme Disease Biobank (LDB)[30,31]. The WB-xVFA achieved 96.9% sensitivity, 96.7% specificity, and 96.8% overall accuracy in blinded testing. These results demonstrate that multiplexed serologic classification can be performed directly from whole blood, establishing a whole-blood-to-answer framework for decentralized Lyme disease testing.

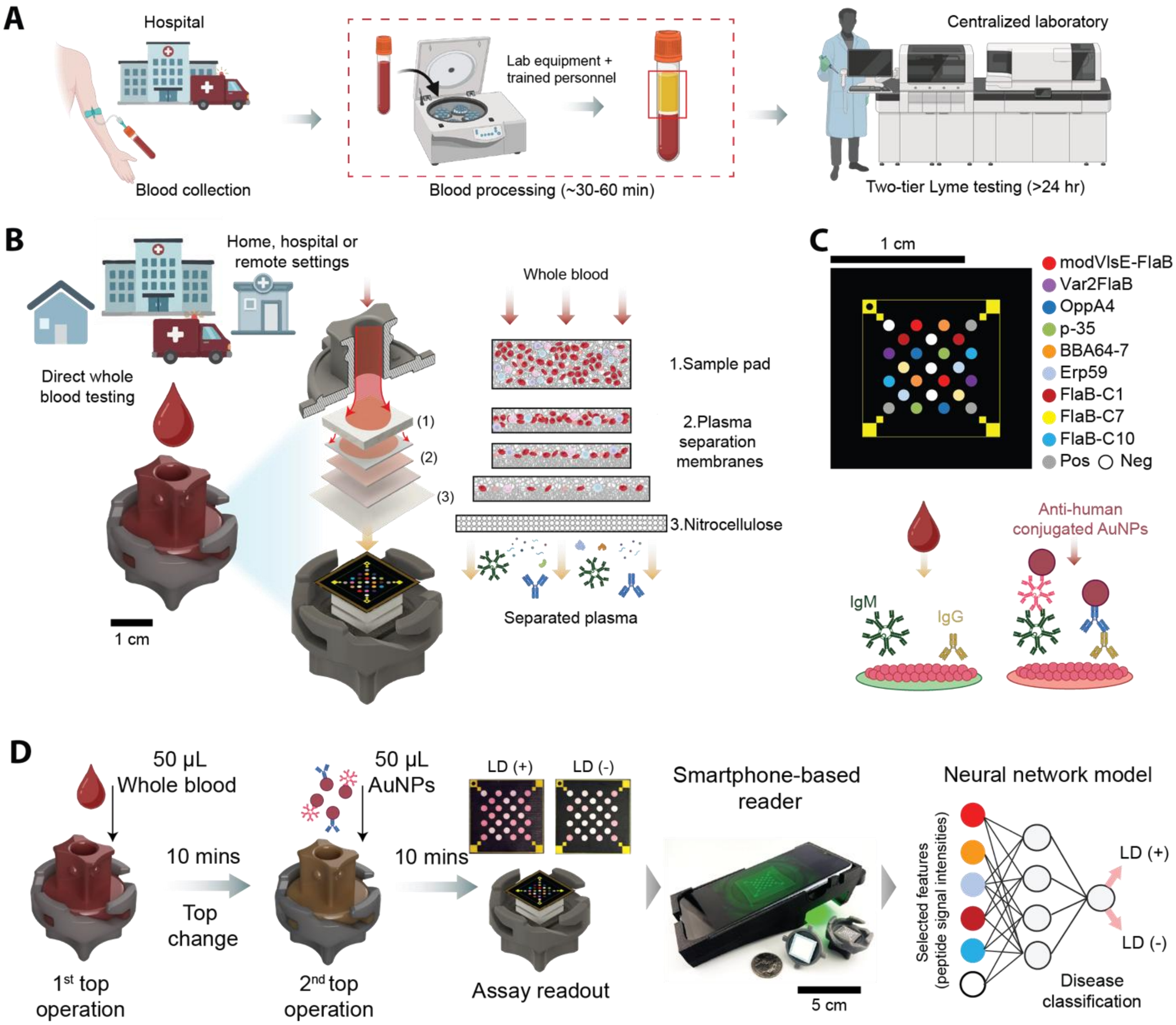

**Fig. 1. Overview of the direct WB-xVFA platform for rapid point-of-care Lyme disease testing using whole blood. (A)** Conventional Lyme disease testing workflow requires venipuncture, laboratory blood processing (~30–60 min; centrifugation and trained personnel), and centralized two-tier serology with a turnaround time of >24 h. **(B)** Schematic of direct whole-blood testing using a stacked WB-xVFA cartridge. Whole blood is applied to the sample pad, and plasma is separated by sequential plasma separation

membranes, enabling a downstream immunoassay without off-device sample processing. **(C)** Representative multiplexed WB-xVFA sensing spot array with immobilized peptides and control spots on the nitrocellulose membrane, including modVlsE-FlaB, Var2FlaB, OppA4, p35, BBA64-7, Erp59, FlaB-C1, FlaB-C7, FlaB-C10, positive control (Pos), and negative control (Neg). Following on-cartridge plasma separation, patient antibodies (IgM/IgG) are captured across the peptide panel and detected using anti-human antibody–conjugated gold nanoparticles (AuNPs), producing spot-wise colorimetric signals. (**D)** Assay operation and readout. A two-step workflow is used: (i) whole blood is applied and allowed to flow through the stacked layers for 10 min, (ii) AuNP labels are added and flowed through to label bound IgM/IgG for 10 min, followed by imaging of the completed array using a smartphone-based reader. Spot intensities are used as features for a neural-network classifier to rapidly output a Lyme disease (LD) positive/negative prediction. Panels (A)–(D) were created with BioRender.com, released under a Creative Commons Attribution-NonCommercial-NoDerivs 4.0 International license (https://creativecommons.org/licenses/by-nc-nd/4.0/deed.en).

## RESULTS

### Design of the WB-xVFA for direct whole-blood testing

Direct whole-blood serology required designing a VFA architecture to process the cellular fraction of whole blood without compromising the flow and spatial signal reproducibility necessary for arrayed multiplexed antibody detection. When whole blood was applied to an initial VFA configuration comprising a series of stacked paper membranes, it produced substantial flow resistance and visible clogging, causing the first sample flow time to exceed the predefined operation-time target of ≤10 min (**Fig. S1**)[21]. Because the membranes are arranged in series, localized obstruction in any layer impedes flow throughout the cartridge. These limitations indicated that whole blood could not be accommodated by simply placing a plasma-separation membrane upstream of the sensing region; rather, the cartridge required an engineered vertical filtration architecture capable of distributing cellular retention across multiple layers to minimize accumulated fluid resistance while maintaining delivery of the antibody-containing fraction to the downstream sensing membrane.

We optimized three key design variables: (i) the effective inlet diameter, (ii) the filtration membrane materials, and (iii) the paper-layer architecture. First, the top-case inlet was redesigned to distribute whole blood across a wider membrane surface, based on the hypothesis that increasing the effective filtration area would reduce localized clogging and improve flow uniformity. We first evaluated a compact inlet geometry compatible with the cartridge footprint; however, the limited effective filtration area led to prolonged whole-blood processing and localized flow resistance. We therefore increased the inlet area to distribute the sample over a larger membrane surface. This change reduced the first-step operation time from 28.0 ± 2.8 min to 9.5 ± 2.1 min, corresponding to a 66% reduction **(Fig. S2)**. However, increasing the inlet area alone was insufficient for robust whole-blood processing, as clogging persisted at higher blood input volumes, indicating that cartridge geometry and filtration-layer design needed to be engineered together.

Using the widened inlet, we next evaluated membrane stacks that varied in material, layer order, and inclusion of a diffuser layer as detailed in **Supplementary Note 1** and **Fig. S3**. Briefly, we investigated plasma-separation membranes and glass-fiber membranes with different wicking rates and blood-filtration capacities to identify materials that could support rapid flow while retaining cellular blood components upstream. Early configurations containing restrictive plasma-separation membranes near the inlet exhibited first-step operation times ranging from 15 to 65 min, substantially exceeding the ≤10-min design target. Among the candidate materials, GR and GX plasma-separation membranes produced shorter operation times (approximately 6 min) and lower mean signal variability (~11%) than the tested glass-fiber membranes, supporting their use in the final graded separation stack (**Fig. S3**). The higher operation times and signal variability observed with the glass-fiber membranes likely reflect localized cellular accumulation and pore obstruction within the fibrous matrix, leading to uneven plasma recovery and downstream delivery. The graded pore structure of the asymmetric GR/GX membranes provided more controlled cellular retention and more consistent transport through the stack[32–34].

We next evaluated layer ordering to determine how the selected materials should be arranged within the vertical stack. Configurations in which a restrictive plasma-separation membrane directly contacted the applied blood were prone to slow flow and clogging. In contrast, positioning a highly porous sample pad upstream promoted lateral spreading of the sample and coarse cellular retention, thereby reducing the filtration burden on the downstream GR and GX membranes. We therefore selected an arrangement in which the upstream sample pad first distributed and partially conditioned the sample before the finer plasma-separation layers completed cellular retention. We also evaluated whether the concentric nitrocellulose diffuser used in previous serum-based xVFA designs was required for whole-blood operation[21,35–37]. Removing the diffuser reduced first-step operation time from 5.4 to 4.7 min, representing an approximately 13% reduction, while maintaining spatially consistent signal development (**Fig. S3**). This result indicated that the widened inlet and upstream sample pad provided sufficient sample distribution without the additional flow resistance introduced by the diffuser.

Based on these design studies, the finalized WB-xVFA architecture consisted of an upstream porous sample pad serving as a coarse pre-filtration and distribution layer, followed by a graded plasma-separation stack comprising a GR membrane, a GX membrane, and a larger GX sealing layer ($1.4 \times 1.4$ cm$^2$), with a downstream 0.22-µm nitrocellulose supporting membrane (**Fig. 2A**). During operation, the sample pad absorbed the whole-blood sample, distributed it across a wider surface area, and partially retained cellular components within its porous fiber network. The subsequent GR/GX plasma-separation layers provided staged removal of remaining cellular components before the antibody-containing fraction reached the sensing membrane. Red blood cells were visibly retained within the upstream filtration layers, whereas the soluble plasma fraction continued through the stack to the sensing membrane (**Fig. 2A**). This on-cartridge filtration enabled whole-blood loading directly into the cartridge while limiting cellular interference with colorimetric immunoreactions and downstream image analysis. Using this architecture, the WB-xVFA achieved consistent and rapid operation across whole-blood samples from multiple donors, with first-step operation time within the ≤10 min design target (**Fig. 2B**). Although Donors 2 and 4 showed a statistically significant difference in operation time following ordinary one-way ANOVA with Tukey's multiple-comparisons test ($p = 0.0334$), all other donors showed no statistically significant difference; this variation was not operationally consequential because all donor samples met the design criterion of ≤10 min and produced positive antibody reactivity. Signal reproducibility across multiplexed peptide spots was strong, with coefficients of variation (CVs) below 5% in Lyme disease patient serum–spiked whole-blood samples, supporting reproducible signal development across the sensing region (**Fig. 2C**). Together, these results demonstrate that the optimized inlet geometry and diffuser-free multilayer filtration stack enabled direct loading of whole blood into the cartridge while preserving the rapid flow and spatial signal reproducibility required for multiplexed serologic analysis.

With the whole-blood filtration architecture established, we implemented the finalized Lyme disease sensing array on the wax-patterned sensing membrane (**Fig. 2D**). The array included replicate spots of nine peptides derived from Lyme-associated antigens: modVlsE-FlaB, Var2FlaB, OppA4, p35, BBA64-7, Erp59, FlaB-C1, FlaB-C7, and FlaB-C10, together with positive and negative control spots, for a total of 25 spots on the test membrane. This layout enabled each whole-blood sample to generate a multiplexed antibody-reactivity pattern for downstream analytical characterization and diagnostic classification.

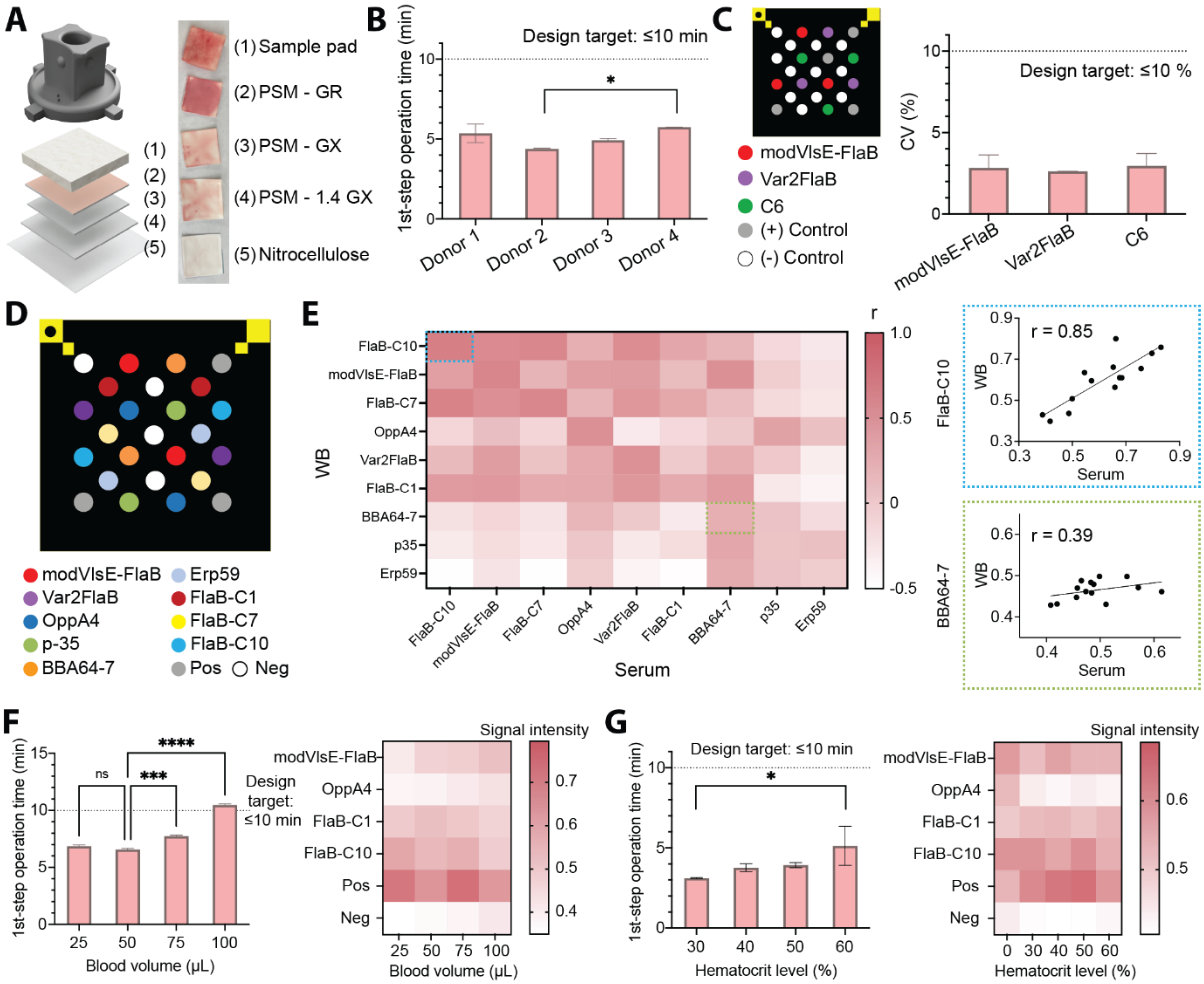

**Fig. 2. Design and analytical characterization of the WB-xVFA. (A)** Exploded schematic of the stacked WB-xVFA cartridge showing the sample pad, sequential plasma separation membrane layers (GR, GX), and nitrocellulose supporting membrane, with representative photographs of each layer taken after assay completion. The visible red coloration in the layers reflects red blood cells retained within the membranes during on-cartridge filtration. **(B)** The first-step operation time is defined as the time required for the top case to wick/process whole blood and the running buffer to traverse the stacked layers (completion of flow through the top-case stack). The dashed line denotes the design target of ≤10 min for rapid assay operation. Data are shown as mean ± SD of duplicate measurements for each whole-blood sample (n = 2). First-step operation time differences among donors were assessed using ordinary one-way ANOVA followed by Tukey's multiple-comparisons test. Donors 2 and 4 differed significantly (p=0.0334); however, all measurements remained below the 10-min design target. Only statistically significant pairwise comparisons are shown. *p<0.05 **(C)** Array layout used for spot-position dependence analysis, the CV (%) calculated

across replicate spots of the same peptide (modVlsE-FlaB, Var2FlaB, C6) printed at different locations. The dashed line denotes the ≤10% design target. Data are shown as mean ± SD of duplicate measurements (n = 2) from one whole-blood sample. **(D)** Spot map of peptide antigens and control spots on the sensing membrane; panels (E)–(G) used the same array layout. **(E)** Cross-correlation heatmap summarizing Pearson correlations (r) between serum and whole-blood signals across the peptide panel, with representative whole-blood-versus-serum scatter plots illustrating a strongly correlated pair (FlaB-C10; r = 0.85) and a weakly correlated pair (BBA64-7; r = 0.39). **(F)** Effect of input blood volume (25–100 μL) on first-step operation time (dashed line denotes ≤10 min design target) and corresponding spot-intensity patterns across selected peptide/control spots (heatmap; color scale at right). Operation-time data are shown as the mean ± SD of duplicate measurements per condition (n = 2) from a single whole-blood sample. Differences among volumes were assessed using ordinary one-way ANOVA followed by Dunnett's multiple-comparisons test, with 50 μL, the standard assay input volume, as the reference condition. Operation time was significantly greater at 75 μL (p=0.0007) and 100 μL (p<0.0001) than at 50 μL. ***p<0.001, and ****p<0.0001 **(G)** Effect of hematocrit level (30–60%) on first-step operation time (dashed line denotes ≤10 min design target) and corresponding spot-intensity patterns across selected peptide/control spots (heatmap; color scale at right). Operation-time data are shown as the mean ± SD of triplicate measurements per condition (n = 3) from a single whole-blood sample. Differences among hematocrit levels were assessed using ordinary one-way ANOVA followed by Tukey's multiple-comparisons test. Operation time differed significantly between 30% and 60% hematocrit (p=0.0182). Only statistically significant pairwise comparisons are shown. *p<0.05.

**Analytical characterization of the WB-xVFA**

Having established a filtration architecture that enabled rapid and reproducible whole-blood operation, we next evaluated whether Lyme-associated multiplexed antibody-reactivity patterns were preserved after transfer from serum into a whole-blood matrix. This direct one-to-one comparison was important because whole blood can introduce matrix effects that alter antibody transport, background signal, and peptide-specific binding behavior, even when cellular components are effectively separated upstream. To assess this transition, fourteen serum specimens (13 Lyme-positive and 1 Lyme-negative) were tested directly and then spiked into presumed Lyme-negative whole blood from a single donor, enabling paired comparisons of serum and whole-blood responses while preserving the same patient antibody profile (**Fig. 2E, Fig. S4**).

Across the peptide panel, Lyme-associated reactivity remained detectable after transfer into the whole-blood matrix, although the degree of agreement varied among individual peptide targets. Same-peptide Pearson correlations ranged from relatively strong agreement for FlaB-C10 (r = 0.85) to weaker agreement for BBA64-7 (r = 0.39) (**Fig. 2E**). These differences indicate that the whole-blood matrix and spiking workflow can differentially affect individual peptide signals. Importantly, these differences did not eliminate the broader panel-level antibody-reactivity pattern, as all 13 paired Lyme-positive serum samples retained Lyme-associated multiplexed reactivity after spiking into whole blood. These results indicate that diagnostically relevant serologic signals were preserved despite peptide-specific matrix effects and support the use of multiplexed computational interpretation rather than reliance on any single peptide feature.

We next evaluated the assay's robustness to practical variations in whole-blood input conditions. Across blood input volumes ranging from 25 to 100 μL, the first-step operation time, defined as the blood-volume-dependent interval required for whole-blood loading and subsequent wash-buffer flow through the filtration stack, increased with sample volume. Relative to the standard 50-μL assay input, operation time was significantly longer at 75 μL (p=0.0007) and 100 μL (p<0.0001) following ordinary one-way ANOVA with Dunnett's multiple-comparisons test. Nevertheless, operation remained below the 10-min design target at 25–75 µL and was only slightly above the target at 100 µL. Corresponding spot-intensity patterns across selected peptide and control features were consistent, indicating that variation in loaded blood volume did not substantially disrupt multiplexed signal generation within the tested range (**Fig. 2F**). We further assessed assay tolerance to hematocrit variation, defined as variation in the fraction of blood volume occupied by red blood cells, because hematocrit differs across patients and can influence whole-blood viscosity, cellular loading, and filtration through porous assay materials. Across hematocrit values spanning 30–60%, first-step operation time was significantly longer at 60% than at 30% hematocrit (p=0.0182, ordinary one-way ANOVA followed by Tukey's multiple-comparisons test). Despite this increase, all measurements remained below the 10-min design target. Spot-intensity patterns across selected sensing locations remained consistent across hematocrit conditions, with most features remaining within approximately ±10–20% of their mean intensity (**Fig. 2G**). These results indicate that increased cellular content did not cause significant deterioration in assay operation within the tested range. Together, the serum-to-whole-blood comparison, volume challenge, and hematocrit challenge demonstrate that the finalized WB-xVFA maintains rapid operation and preserves diagnostically relevant multiplexed antibody-reactivity patterns across practical whole-blood testing conditions.

**Single-tier WB-xVFA Lyme disease diagnostic algorithm development and blind testing**

Using the finalized assay workflow detailed in the former section, we tested 60 frozen whole-blood samples from the LDB (30 laboratory-confirmed Lyme disease, 30 endemic controls) in duplicate, generating 120 WB-xVFA measurements. Because blinded evaluation relied on frozen whole-blood specimens from the LDB, the wash-buffer formulation was optimized before clinical testing to improve signal uniformity after freeze–thaw processing (Fig. S5–S7). The optimized formulation was incorporated into the finalized WB-xVFA workflow used for all subsequent clinical evaluations. In addition, 16 serum samples (14 Lyme-positive, 2 Lyme-negative) were tested in duplicate to generate 32 internal-control measurements that were used exclusively for model training. These serum-derived measurements were included as internal controls to expand the diversity of antibody-reactivity patterns available during model development and were not used for blinded evaluation. The full dataset comprised 152 WB-xVFA measurements and was partitioned into a model-development cohort (n = 90 tests) and an independent, blinded-evaluation cohort (n = 62 tests), with no overlap between the two sets. Duplicate measurements were kept within the same partition to prevent information leakage between model development and blinded testing (**Fig. 3A**). We first examined the multiplexed signal patterns from the training cohort. **Fig. 3B** shows the peptide spot map used for the clinical study, together with normalized WB-xVFA signal intensities across the full Lyme peptide and control panel. Lyme-positive tests generally showed stronger reactivity across multiple peptide features than Lyme-negative tests, as desired, but the magnitude and distribution of signal varied across samples and antigens. This heterogeneous response pattern indicated that disease-associated information was distributed across the peptide panel rather than concentrated in any single marker, motivating multiplexed computational analysis, rather than relying on thresholding a single biomarker signal.

To convert the multiplexed WB-xVFA reactivity patterns into a binary diagnostic output, we developed a neural network classifier trained solely on the training cohort. Background-normalized spot intensities ($x_i^j$) from the immunoreaction array were used as candidate input features, and model development included both hyperparameter optimization and input-feature selection. This framework allowed machine learning to serve two roles: selecting the most informative sensing features and performing automated diagnostic classification. The reduced feature set may also enable simplification of the sensing-spot array, thereby helping establish an optimal trade-off between assay cost, complexity, and performance. Through this process, the compact classifier was selected as a shallow fully connected neural network with two hidden layers; see the Methods for details. At the output layer, the model contained a single unit with a sigmoid activation function, providing a prediction score in the [0, 1] range, which was compared against a 0.5 threshold to perform the final binary Lyme disease diagnostic decision; ≥0.5 classified as Lyme positive and <0.5 as Lyme negative.

In our WB-xVFA system, input features were selected via an iterative sequential forward feature selection (SFFS) process, in which individual immunoreaction spots were added to the model one at a time[19,38]. At each iteration, the candidate spot that produced the highest cross-validated area under the receiver operating characteristic (AUC-ROC) when added to the current feature set was retained for the next iteration, until all 25 spots were included in the panel. We intentionally performed feature selection at the individual spot level rather than relying solely on peptide-averaged intensities to preserve the spatial information of the spot arrangement on the membrane as an optimizable feature. The feature set yielding the highest AUC was selected for the final model configuration (**Fig. 3B, C**). Feature optimization converged to a six-feature subset comprising five Lyme-associated peptide spots (FlaB-C10, modVlsE-FlaB, FlaB-C1, Erp59, and BBA64-7) and one negative-control spot. These selected peptides represent complementary *Borrelia burgdorferi* antibody targets spanning FlaB-, VlsE-, OspC-, Erp59-, and BBA64-associated epitopes, indicating that diagnostic discrimination relied on an integrated antibody-reactivity pattern rather than any single antigen response. The negative-control feature likely helped capture nonspecific background variations across tests.

With this six-feature input, the optimized shallow neural network achieved an AUC of 1.0 on the training set, with 98.9% accuracy and only one false-positive prediction among the duplicate assay measurements (**Fig. 3D**). The false-positive result occurred in one of the two independent WB-xVFA runs from the same biological sample, while the second run was classified correctly. The finalized six-feature neural network classifier was trained on the complete training dataset, locked, and then evaluated on an independent blinded cohort of unseen patient samples.

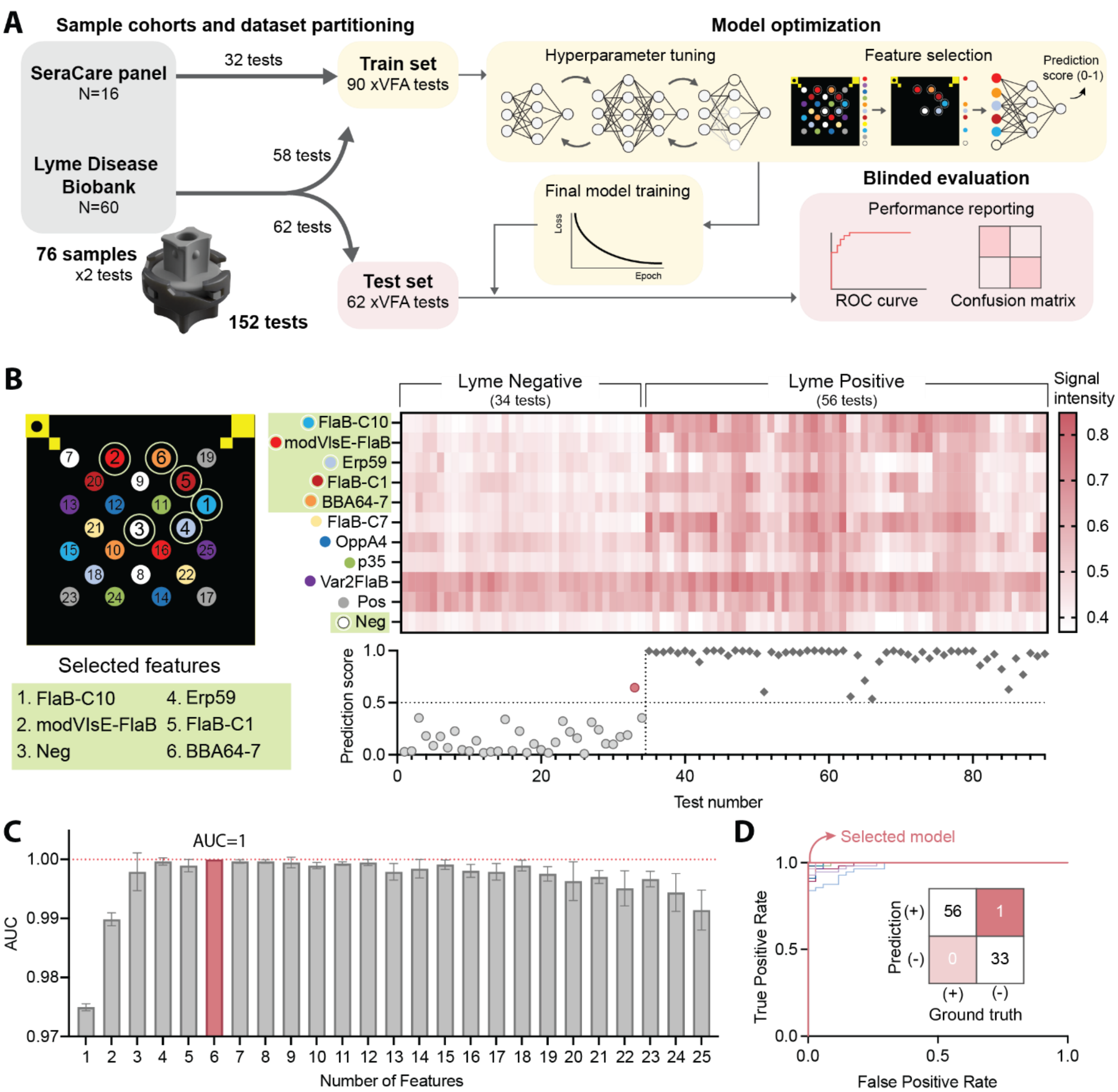

**Fig. 3. Training and feature selection of the shallow neural network-based classifier for whole-blood Lyme disease testing. (A)** Clinical cohort composition, duplicate test handling, and data-splitting strategy for neural-network model development. Clinical samples were obtained from two sources (SeraCare panel, N = 16; Lyme disease biobank, N = 60; 76 total samples) and tested in duplicate on the WB-xVFA (152 total tests). Tests were partitioned into a training set (90 WB-xVFA tests) and an independent blind testing set (62 WB-xVFA tests). Duplicate tests from the same sample were kept within the same partition and were treated as separate tests. Model optimization on the training set included hyperparameter tuning and feature selection, followed by final model training and blinded evaluation on the held-out test set. **(B)** Spot map of peptide antigens and control spots with the optimized feature spots for the neural network input (FlaB-C10, modVlsE-FlaB, Erp59, FlaB-C1, BBA64-7, negative-control) highlighted in green and a heatmap of normalized spot intensities (averaged across replicates) for the training set, grouped by Lyme-negative (34 tests) and Lyme-positive (56 tests). The bottom panel shows neural network prediction scores (0–1) for the training set; scores ≥0.5 were classified as Lyme-positive and scores <0.5 as Lyme-negative. The vertical dashed line separates Lyme-negative (left) and Lyme-positive (right) tests along the x-axis. **(C)**

AUC for each iteration of the feature-selection process; the selected feature set (6 features) is highlighted with green circles in panel (B). **(D)** ROC curves for candidate feature sets with the selected feature set highlighted; inset confusion matrix summarizes training-set performance for the selected model (n = 90 tests).

To test whether the finalized neural network-based classifier generalized beyond the model-development sample cohort, we evaluated it on an independent, blinded set of LDB specimens with verified two-tier serology reference labels. This cohort included 31 unique clinical specimens (15 Lyme-negative and 16 Lyme-positive) tested in duplicate, yielding 62 WB-xVFAs: 30 Lyme-negative and 32 Lyme-positive tests. None of these specimens were used during feature selection, model optimization, or training, allowing us to assess WB-xVFA performance on previously unseen clinical whole-blood samples.

In the blinded cohort, normalized WB-xVFA signal heatmaps showed distinct group-level reactivity patterns between Lyme-negative and Lyme-positive tests, with Lyme-positive measurements generally exhibiting stronger responses across Lyme-associated peptide features (**Fig. 4A**). The corresponding prediction-score plot showed that all Lyme-positive tests except one had scores above the fixed 0.5 decision threshold, while one Lyme-negative test also had a score above the threshold. The false-negative and false-positive results each occurred in one of the two WB-xVFA runs from a single specimen, and the other run from the same specimen was classified correctly. Across the 62 blinded tests, the classifier produced 31 true positives, 29 true negatives, 1 false negative and 1 false positive, yielding 96.8% accuracy, 96.9% sensitivity and 96.7% specificity (**Fig. 4B**).

To evaluate whether this performance depends on the selected modeling strategy, we benchmarked the finalized shallow neural network-based classifier against alternative machine learning approaches, including logistic regression and random forest models (**Fig. S8**). To enable a fair comparison, all models were trained and blindly evaluated using the identical training and testing partitions. Because each learning algorithm may require a different feature representation, logistic regression and random forest models underwent independent SFFS optimization before model training, yielding optimal feature sets of four and three features, respectively. On the blinded test set, logistic regression achieved 91.9% accuracy, and random forest achieved 82.3% accuracy, both inferior to the neural network classifier results reported earlier. Furthermore, an alternative neural network trained on all 25 immunoreaction spots also showed reduced blind-testing performance, with 87.1% accuracy, 93.3% sensitivity, and 81.2% specificity, indicating that feature selection improved the overall disease classification performance by reducing the influence of weakly informative and potentially noisy inputs.

It is also important to note that single-peptide thresholding across the nine Lyme-associated peptide features produced accuracies ranging from 53% to 92%, and no individual peptide-averaged signal exceeded the performance of the six-feature neural network (see Supplementary Note 2). Together, these comparisons indicate that the optimal blinded performance of the WB-xVFA was driven by feature-selected multiplexed interpretation through a shallow neural network-based classifier, rather than by a single dominant antigen response or by an unoptimized full-panel model.

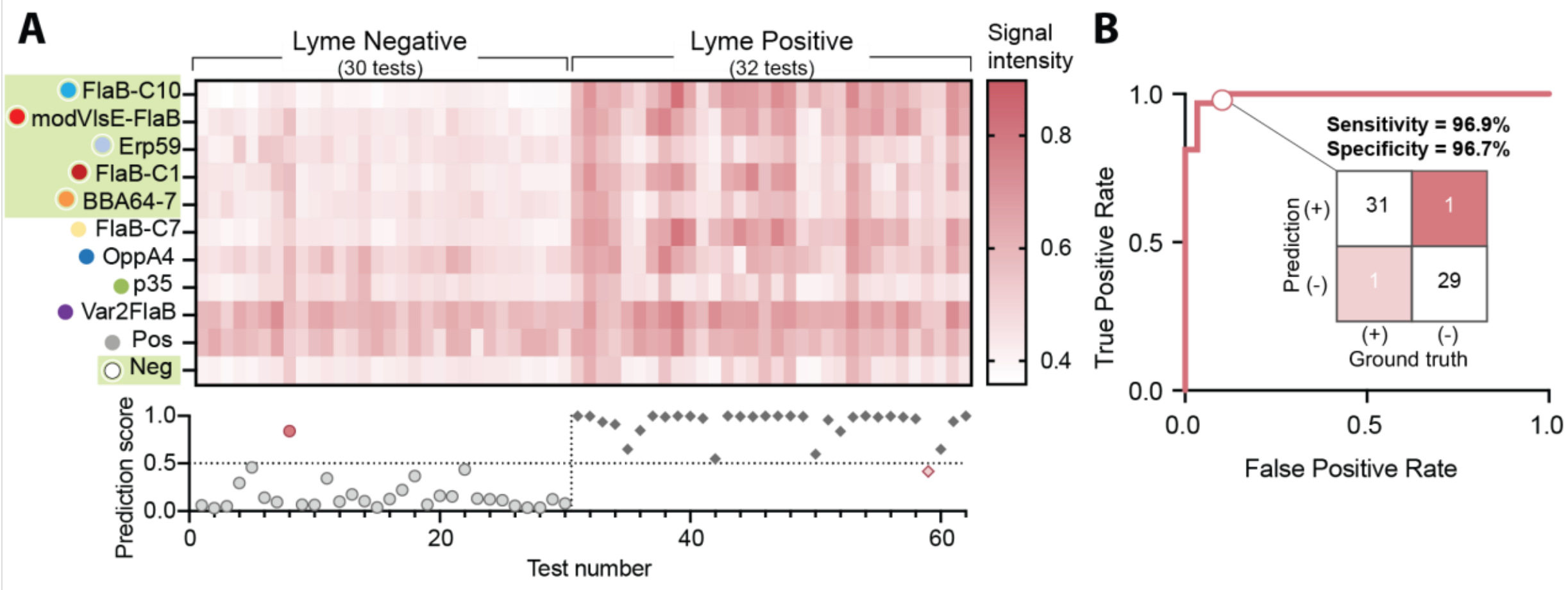

**Fig. 4. Blinded evaluation of the shallow neural network-based classifier. (A)** Blinded evaluation of the WB-xVFA with the optimized shallow neural-network classifier on 31 clinical samples (15 Lyme-negative, 16 Lyme-positive) from the LDB, tested in duplicate (62 blinded tests; 30 Lyme-negative and 32 Lyme-positive). Top, heatmap of normalized spot-intensities (averaged across replicates) for the blinded set. Bottom, model prediction scores (0–1) for each blinded test, with the classification threshold at 0.5 (horizontal dashed line); circles denote Lyme-negative tests and diamonds denote Lyme-positive tests; the vertical dashed line separates the two groups along the x-axis. **(B)** ROC curve and confusion matrix for the 62 blinded tests from 31 clinical samples, corresponding to a sensitivity of 96.9% (31/32) and a specificity of 96.7% (29/30). Duplicate measurements from the same sample were treated as separate tests.

## DISCUSSION

This study establishes a whole-blood workflow for rapid single-tier Lyme disease serology by integrating on-cartridge blood processing, multiplexed peptide-based antibody detection, portable reader, and neural-network interpretation into a rapid point-of-care platform. By incorporating whole-blood preparation and serological analysis in a single test cartridge, the need for off-device clotting, centrifugation, serum or plasma transfer, and sequential laboratory testing was eliminated. These upstream requirements remain a major barrier to decentralized serology because they increase processing time, manual handling, and dependence on laboratory infrastructure and trained personnel. By accepting whole blood directly and producing an objective result within 20 min, the WB-xVFA provides a whole-blood-to-answer architecture for near-patient Lyme disease testing.

A central technical contribution of this work is the demonstration that whole-blood processing and multiplexed serologic sensing can be co-designed within a single paper-based diagnostic architecture. Direct whole-blood serology required more than placing a conventional plasma-separation membrane upstream of an existing immunoassay. Whole blood introduces coupled challenges, including increased flow resistance, cellular obstruction, matrix-dependent signal variation, and interference with colorimetric readout. Blood processing and serologic sensing, therefore, had to be engineered as an integrated system in which efficient cellular removal, antibody transport, and multiplexed signal generation were jointly optimized. This required coordinated changes to the inlet geometry, filtration materials, membrane order, and assay chemistry. The enlarged inlet distributed the sample across a wider filtration area, while the graded multilayer stack distributed cellular retention across successive materials rather than concentrating it at a single restrictive interface. This architecture also achieved uniform flow without a concentric diffuser layer, indicating that the enlarged inlet and multilayer filtration stack provided sufficient sample distribution while avoiding additional flow resistance. The resulting cartridge maintained rapid flow and reproducible multiplexed signal development across the tested whole-blood conditions. These findings show that direct whole-blood compatibility arose from coordinated design of the sample-processing and sensing functions rather than from simply substituting whole blood for serum.

The study also illustrates how clinical specimen format can influence the translation of whole-blood assays. Blinded evaluation required frozen biobanked specimens, which represent a more challenging matrix than freshly collected whole blood because freeze–thaw processing can induce hemolysis and broader physical and biochemical changes in the blood matrix. These changes, including the release of intracellular components and altered sample composition, can affect fluid transport, nonspecific interactions, and signal development within porous diagnostic materials. The greater spatial signal non-uniformity observed with frozen specimens therefore motivated optimization of the assay chemistry for the specimen format used in clinical evaluation. Incorporation of Triton X-100 into the wash buffer improved signal uniformity across both fresh and frozen whole-blood conditions, enabling the finalized workflow to support blinded testing of clinical specimens.

Matrix-comparison experiments demonstrated that the transition from serum to whole blood also altered the magnitude of individual peptide responses while preserving the overall multiplexed antibody-reactivity profile. Individual peptide signals displayed different levels of agreement between serum and matched serum-spiked whole blood, indicating that matrix-dependent transport, background and binding effects influenced signal intensity across specific assay features. Despite these matrix-dependent effects, the assay maintained diagnostically relevant performance under practical whole-blood testing conditions. The consistent signal patterns observed across blood volumes from 25 to 100 μL and hematocrit values from 30% to 60% further support the vertical-flow architecture's ability to tolerate practical variation in whole-blood composition and loading.

This work establishes a foundation for prospective clinical translation of direct whole-blood Lyme disease serology. In a blinded evaluation of 31 unique clinical specimens tested in duplicate, the WB-xVFA demonstrated robust analytical and diagnostic performance, supporting the feasibility of integrating blood processing and serological analysis within a single rapid assay. Future prospective, multisite studies using the WB-xVFA with freshly collected capillary whole blood will be important for validating performance across broader patient populations, disease stages, operators, and clinical settings representative of intended point-of-care use. These studies should also evaluate analytical specificity in larger cohorts with potentially cross-reactive infections, autoimmune or inflammatory conditions, together with cartridge manufacturing reproducibility, reagent stability, lot-to-lot consistency, and decentralized usability.

Overall, the WB-xVFA demonstrates that rapid multiplexed peptide serology can be performed directly from whole blood through the integrated co-design of blood processing, multiplexed peptide sensing, and machine-learning interpretation within a single assay. By eliminating off-device sample preparation while preserving diagnostically informative multiplexed antibody-reactivity patterns, the platform provides a practical approach to decentralized Lyme disease serology. More broadly, this work establishes a general engineering strategy for translating multiplexed serological assays from processed serum or plasma to whole blood. This framework may enable the development of future whole-blood multiplexed serological assays for Lyme disease and other infectious diseases requiring rapid near-patient antibody testing.

## METHODS

### Ethics statement

This study used de-identified and blinded human serum and whole-blood specimens obtained from Lyme Disease Biobank. Patient specimens acquired from the Lyme Disease Biobank were collected with informed consent under Institutional Review Board approval through Advarra IRB protocol Pro00012408. Because only de-identified specimens were provided to UCLA investigators and no identifiable participant information was available to the study team, the research conducted at UCLA did not require UCLA IRB review. Commercial SeraCare specimens were obtained as de-identified reference materials and used in accordance with the supplier's terms and applicable institutional requirements. No identifiable participant information was available to the study investigators.

### Study design, clinical specimens, and dataset partitioning

Two specimen cohorts were used for model development and blinded evaluation. The first cohort consisted of a commercial Lyme disease performance panel from SeraCare, comprising 16 serum specimens, including 14 Lyme-positive and 2 Lyme-negative samples. The second cohort consisted of 60 frozen whole-blood specimens with confirmed standard two-tier testing results from the Lyme Disease Biobank, including 30 Lyme disease samples with positive results and 30 seronegative endemic controls. All specimens were tested in duplicate using the WB-xVFA platform. Duplicate measurements from the same specimen were kept within the same data partition to prevent information leakage between model development and blinded evaluation.

The complete dataset comprised 152 WB-xVFA measurements, including 32 measurements from the SeraCare serum panel and 120 measurements from the Lyme Disease Biobank whole-blood specimens. All SeraCare measurements and 58 measurements from 29 Lyme Disease Biobank specimens were assigned to the model-development cohort, yielding 90 WB-xVFA tests for training, hyperparameter optimization, and feature selection. The remaining 31 Lyme Disease Biobank specimens were reserved as an independent blinded-evaluation cohort and tested in duplicate, yielding 62 WB-xVFA measurements. The final classifier was trained only on the model-development cohort and locked before evaluation on the blinded cohort.

### WB-xVFA assay design

The WB-xVFA consists of a two-part plastic cassette that houses a multilayer filtration stack and a multiplexed nitrocellulose sensing membrane. The sensing membrane was prepared using the hydrophobic patterning and peptide-spotting method described previously[21]. Briefly, 25 hydrophobically isolated sensing regions were defined by wax printing and functionalized with Lyme-associated peptide antigens, together with internal positive-control spots (goat anti-mouse IgG; SouthernBiotech, 1036-01) and negative-control spots (1% BSA in PBS, pH 7.4). The antigen panel included peptides used in the serum-based xVFA together with newly developed dual-epitope peptides designed to improve epitope specificity and diagnostic performance. The first top case contains a vertical filtration stack designed to retain blood cells during capillary-driven flow while allowing the antibody-containing plasma fraction to flow through to the sensing membrane. The first top case, therefore, supports whole-blood filtration and delivery of the separated plasma fraction to the sensing membrane. After completion of the first assay step, the top case is replaced with a second top case that delivers gold nanoparticle (AuNP)-conjugated anti-human IgM/IgG detector antibodies for colorimetric signal development. The AuNP detector conjugate was prepared using the previously reported method[19,21]. Pre- and post-assay images of the sensing membrane were acquired using a smartphone-based reader with uniform green-LED illumination and fixed imaging geometry.

**WB-xVFA assay operation**

Before sample loading, a background image of the unused sensing membrane is captured with the smartphone reader. The cassette is then assembled by twisting the first top case onto the bottom case. In the first top-case step, 200 µL of running buffer, 50 µL of whole blood, and an additional 200 µL of running buffer are sequentially added to the sample inlet. The initial running buffer aliquot pre-wets and activates the membrane assembly, whereas the final aliquot facilitates continued flow through the filtration stack and promotes removal of unbound sample components. During this step, RBCs and white blood cells are retained within the filtration layers while plasma flows to the sensing membrane. After the reagents are absorbed into the cassette, typically within ~10 min, the first top case is removed and replaced with the second top case for signal generation. In the second step, 300 µL of running buffer is first added to pre-wet and activate the membrane assembly and to help clear residual unbound blood-derived material from the sensing membrane. This is followed by 50 µL of AuNP-anti-human IgM/IgG detector solution and 200 µL of running buffer. The detector mixture flows through the membrane for approximately 10 min, labeling bound Borrelia-specific antibodies while excess conjugate is removed by the running buffer. At completion, the cassette is opened, and the sensing membrane is imaged. Normalized spot intensities are computed by normalizing post-assay values to the corresponding pre-assay background signals, as described below.

**Smartphone-based reader and image processing**

Pre-assay (background) and post-assay (signal) images of the sensing membrane are acquired using the smartphone-based reader under fixed illumination and exposure settings. The reader consists of a smartphone (LG G7 ThinQ) integrated with a custom 3D-printed attachment, printed using a Stratasys 3D printer. This attachment includes an optical module consisting of 4 green LEDs (525 nm), an external lens, an LED driver circuit to provide a stable 20 mA current, and a power jack. The LED peak wavelength was selected to match the absorption peak of the AuNPs, which stays in the 520–530 nm range. The LEDs are arranged in a circular pattern around the external lens and polished from the front to provide uniform illumination of the sensing membrane. A custom 3D-printed tray is used to insert WB-xVFA cartridges into the reader, providing a dark background and a fixed assay position within the reader's field of view to minimize inter-sensor and inter-user variations. The reader captures images of the sensing membrane under fixed illumination and exposure settings (~0.9 ms exposure, ISO 50) in raw DNG format. These images are transferred to a desktop computer, converted to TIFF format, and the green channel is extracted to maximize contrast generated by AuNP absorption.

All 25 immunoreaction spots per test are automatically segmented from the smartphone image using a custom segmentation code (in Python), where circular masks covering approximately 80% of the spot area are overlayed with each spot. Final spot intensities are computed as the mean pixel value within each mask and normalized by dividing post-assay signals ($s_i^j$) by the corresponding background values ($b_i^j$) using the following equation:

$$x_i^j = 1 - \frac{s_i^j}{b_i^j},$$

where $i$ is the type of the immunoreaction spot (i.e., $i \in$ {modVlsE-FlaB, Var2FlaB, OppA4, p35, BBA64-7, Erp59, FlaB-C1, FlaB-C7, FlaB-C10, positive control, negative control}) and $j$ is the spot repeat (i.e., $j$ = 1-2 for peptide spots, $j$ = 1-3 for positive control spots, and $j$ = 1-4 for negative control spots). These $x_i^j$ values are used for all downstream analyses, including optimization experiments and neural network-based analysis.

**Optimization of whole-blood filtration conditions**

To adapt the vertical flow cartridge for direct whole-blood testing, we evaluated the filtration design in a stepwise manner using donor whole blood. Optimization focused on three design parameters: inlet geometry, membrane selection, and membrane-stack architecture. First-step operation time was defined as the sum of the whole-blood wicking time and the subsequent running-buffer wicking time. Each interval was measured manually using a stopwatch, beginning immediately after liquid addition and ending when no visible liquid remained at the top-case inlet. The two measured intervals were summed to obtain the total first-step operation time. A design target of ≤10 min was used during assay development.

We first evaluated the effect of inlet geometry on whole-blood flow. Two top-case inlet designs were compared: the original narrow inlet and an enlarged inlet with increased effective inlet diameter. For this comparison, 40 µL of donor whole blood was applied to each inlet design, and the first-step operation time was recorded (**Fig. 2 and S2**). Using the enlarged inlet geometry, we next investigated candidate filtration materials with different wicking rates and blood-filtration capabilities. Candidate membranes were assembled into top-case stacks and tested using donor whole blood to identify materials that supported rapid vertical flow while retaining cellular blood components upstream. Initial architecture-screening experiments were performed using 60 µL of donor whole blood. Follow-up comparisons of refined stack architectures were performed using 50 µL of donor whole blood.

We then evaluated membrane-layer ordering to determine how the selected materials should be arranged within the vertical stack. Alternative stack configurations were constructed by varying the order of the top two layers, with either the sample pad or the plasma-separation membrane serving as the first-contact layer for the applied whole blood. These experiments were performed using 50 µL of donor whole blood. Performance was assessed by success rate, first-step operation time, and average signal variability. Success rate was defined as the fraction of tests (n = 5) in which the applied sample and running buffer were fully processed without visible clogging or residual liquid at the top-case inlet. Average signal variability was calculated as the mean coefficient of variation (CV) across the C6, positive-control, and negative-control features, with each feature-level CV calculated from spatially replicated spots of the same feature on the sensing membrane. Finally, we evaluated whether the diffuser layer used in prior serum-based xVFA architectures was required for whole-blood operation. Otherwise identical stack configurations were assembled with or without the diffuser layer and tested using 50 µL of donor whole blood. Plasma-separation membrane materials were also compared within a common stack architecture by substituting candidate membranes while keeping the remaining layers unchanged. For these comparisons, first-step operation time and cumulative signal variability were evaluated using the criteria described above (**Fig. S3**).

**Analytical Characterization of the WB-xVFA**

Analytical characterization of the finalized WB-xVFA included comparisons of serum and whole-blood matrix responses, evaluation of robustness across blood input volumes, and assessment of hematocrit compatibility. For serum–whole-blood matrix comparison, 14 serum specimens from SeraCare were tested both directly (20 µL serum) and after spiking the same 20 µL serum aliquots into 50 µL of presumed-negative donor whole blood. Peptide-wise signal correspondence between the serum and serum-in-whole-blood conditions was then assessed by Pearson correlation calculated across matched serum and serum-in-whole-blood measurements for each peptide. For blood-volume robustness testing, whole-blood input volumes of 25, 50, 75, and 100 µL were evaluated using the first-step operation time and qualitative assessment of resulting spot-intensity patterns. For hematocrit tolerance testing, whole-blood samples were adjusted to nominal hematocrit values of 30, 40, 50, and 60% by centrifuging whole blood from a single donor, separating plasma from the packed RBC fraction, and recombining measured plasma volumes with the packed RBC fraction to generate the target hematocrits. Because hematocrit values were defined by volumetric recombination rather than independently measured after recombination, they are reported as nominal values. These samples were then processed with the finalized assay configuration, and hematocrit tolerance was assessed by the first-step operation time and qualitative spot-pattern evaluation.

**Optimization of the wash-buffer of WB-xVFA**

The intended end-use format of the WB-xVFA is freshly collected whole blood. However, blinded clinical evaluation was performed using frozen whole-blood specimens from the Lyme Disease Biobank (LDB). To evaluate the compatibility of the WB-xVFA with frozen whole-blood specimens, three different Lyme-positive serum specimens were individually spiked into whole blood from three different presumed-negative donors and tested before freezing and after one freeze–thaw cycle (4-day storage at −80 °C). Each test used 50 µL of whole blood (**Fig. S5**). Signal uniformity was quantified by comparing normalized intensities between the central region (zone a) and outer region (zone b) of the sensing membrane. Statistical analysis was performed using ordinary two-way ANOVA with Šídák's multiple-comparisons test to compare zone a and zone b within each condition (**Fig. S6**). For fresh-versus-frozen whole-blood compatibility, 15 Lyme-positive serum specimens were spiked into presumed-negative donor whole blood and tested under the finalized assay conditions before freezing and after one freeze–thaw cycle (4-day storage at −80 °C). One specimen failed assay quality control and was excluded from downstream analysis, resulting in 14 paired fresh and frozen serum-spiked whole-blood measurements (**Fig. S7**).

**Training the single-tier WB-xVFA neural network model**

A shallow, fully connected neural network-based classifier was used for binary diagnosis of Lyme disease. The input layer of this classifier incorporated nodes corresponding to the background-normalized spot signals ($x_i^j$). The neural network model consisted of two hidden layers (128, 32 units), each with rectified linear unit (ReLU) nonlinear activation functions, L2 regularization ($\lambda$ = 1e-4), batch normalization, and 50% dropout. The model was trained using the binary cross-entropy loss function and the Adam optimizer (learning rate = 1e-3, batch size = 5). Hyperparameters, including network depth, number of units per layer, dropout rate, regularization parameter ($\lambda$), learning rate, and batch size, were optimized via grid search using the training/validation dataset only, i.e., test samples were excluded. The binary cross-entropy loss function ($L_{BCE}$) is defined as:

$$L_{BCE} = -\frac{1}{N_b}\sum_{n=1}^{N_b} y_n \log y_n' + (1 - y_n)\log(1 - y_n'),$$

where $y_n$ represent ground-truth Lyme disease diagnostics labels (i.e., $y_n \in \{0,1\}$), $N_b$ is the batch size, and $y_n'$ denotes the model's predicted probability score in the [0, 1] range, based on the output from the sigmoid activation function defined as:

$$y_n' = \frac{1}{1 + e^{-\hat{y}_n}},$$

Here, $\hat{y}_n$ is the model's inference before the sigmoid function. Samples with output scores above or equal to 0.5 were assigned to Lyme-positive, while samples with the score below the 0.5 threshold were assigned to Lyme-negative.

In addition to optimizing hyperparameters, we also optimized the input features of the diagnostic model using the iterative SFFS process (see the "Feature selection using sequential forward feature selection (SFFS)" for more details). As a result of this procedure, the model's input incorporated 6 input features (out of 25), including 5 peptide spots and 1 negative-control spot. Both hyperparameter and input feature optimizations were performed using a 4-fold cross-validation on the training/validation set, excluding test samples. The optimized model (including optimal hyperparameters and input features) achieved 98.9% accuracy, 100% sensitivity, and 97.1% specificity on the validation set. The optimized model was further trained on the full training/validation dataset and blindly evaluated on a separate test set, consisting of samples not seen by the model during the optimization and training stages. On the blind testing set, the diagnostic model achieved 96.8% accuracy, 96.9% sensitivity, and 96.7% specificity, as reported in the "Results" section. See the "Study design, clinical specimens, and dataset partitioning" section for details on the sample partitioning between training/validation and blind testing sets. The neural network model was trained and optimized on a desktop computer equipped with a GeForce GTX 1080 Ti graphics processing unit (NVIDIA). The model was created in Python using NumPy, TensorFlow, and Keras libraries. Training time for the classifier model was ~15 s, while blind testing time was substantially shorter, taking less than 0.3 s to process all 62 blind-testing samples from the LDB.

**Feature selection using sequential forward feature selection (SFFS)**

SFFS was performed on the training/validation set to identify an optimal subset of immunoreaction spots for the neural network-based classifier. At each iteration, one candidate spot-level feature was added to the neural network's input layer, the model was retrained via 4-fold cross-validation, and performance was reevaluated using AUC. The feature that produced the greatest improvement in AUC was retained, and the process continued until all 25-candidate spot-level features, including positive and negative controls, had been incorporated. The resulting optimal feature subset identified by SFFS included 6 spots: FlaB-C10, modVlsE-FlaB, FlaB-C1, Erp59, BBA64-7 peptide spots and a negative-control spot. This subset was used for all subsequent neural network training and evaluation on the blind testing set. To provide a fair comparison, feature selection was performed independently for other machine learning classifiers reported in the Results section, including logistic regression and random forest, following the SFFS procedure described above.

**Statistical analysis**

Statistical analyses were performed in Python using OpenCV, NumPy, pandas, SciPy, scikit-learn, TensorFlow/Keras, and Matplotlib. $x_i^j$ values were used for all quantitative analyses. Signal variability across replicate spots was quantified using the CV (%), defined as the standard deviation divided by the mean signal value. For filtration-optimization experiments, first-step operation time, success rate, and cumulative signal variability were used as performance metrics. Success rate was defined as the fraction of tests in which whole blood fully traversed the top-case stack without visible clogging. Average signal variability was calculated as the mean of CV values across selected peptide and control features.

Differences in first-step operation time among whole-blood donors were assessed using ordinary one-way ANOVA followed by Tukey's multiple-comparisons test. Differences in first-step operation time among blood-input volumes were assessed using ordinary one-way ANOVA followed by Dunnett's multiple-comparisons test, with 50 µL, the standard assay input volume, specified as the reference condition. Differences in first-step operation time among hematocrit conditions were assessed using ordinary one-way ANOVA followed by Tukey's multiple-comparisons test. Only statistically significant pairwise comparisons are shown in the figures. Statistical significance was defined as $p < 0.05$.

For serum-to-whole-blood comparisons, peptide-wise agreement between matched measurements was assessed using Pearson correlation coefficients (**Fig. 2E**). Spatial signal-uniformity analyses compared normalized spot intensities between defined sensing-membrane regions. Triton X-100 optimization was analyzed using ordinary two-way ANOVA with Šídák's multiple-comparisons test (**Fig. S6**). For membrane–diffuser uniformity experiments, spot signals in the inside and outside zones were compared using a two-sample t-test, with lower absolute t-scores indicating more uniform reagent delivery.

Diagnostic performance was quantified using receiver operating characteristic curves, the area under the curve, accuracy, sensitivity, specificity, and confusion matrices. Sensitivity was defined as true positives divided by true positives plus false negatives, specificity as true negatives divided by true negatives plus false positives, and accuracy as true positives plus true negatives divided by the total number of tests. Duplicate measurements from the same specimen were kept within the same data partition to prevent information leakage between model development and blinded evaluation.

## ACKNOWLEDGEMENTS

The authors acknowledge the Lyme Disease Biobank at the Bay Area Lyme Foundation for providing clinical whole-blood samples essential for validating this platform.

## FUNDING

This work was supported by the National Institutes of Health (NIH) (Grant R44AI150060) awarded to P.M.A. and D.D.C.; the NSF Partnerships for Innovation (PFI-TT) program (Award ID: 2345816) awarded to A.O. and D.D.C.; and the National Science Foundation (NSF) PATHS-UP Engineering Research Center (Grant 1648451) awarded to A.O. and D.D.C. B.P. acknowledges support from the NSF Graduate Research Fellowship Program (GRFP) under Grant No. DGE-2034835. Schematic figures and illustrations were partially created using BioRender.com.